\newif\ifrevtex
\newif\ifsubmission
\newif\iftwocolumn

\revtextrue   

\submissionfalse \twocolumntrue   

\ifrevtex\else\submissionfalse\twocolumntrue\fi

\ifrevtex
  \ifsubmission
    \documentstyle[floats,prl,aps,preprint,times]{revtex}
    \twocolumnfalse
  \else
   \documentstyle[floats,prl,aps,multicol,times,epsf]{revtex}
   \tighten
  \fi

  \renewcommand\ensuremath\relax
  \newcommand\eqref[1]{(\ref{#1})}

  \newcommand\citep[1]{\cite{#1}}
  \renewcommand{\usepackage}{\RequirePackage}
  \usepackage{times,graphicx,psfrag,subfigure,url}
\else
  \documentclass[draft,prb,aps,twocolumn,tightenlines]{revtex4}
  \usepackage{times,psfrag,amsmath,subfigure}
  \usepackage[draft]{graphicx}
  \DeclareMathAlphabet{\mathrmb}{OT1}{ptm}{b}{n}
  \DeclareMathAlphabet{\mathsfb}{OT1}{phv}{b}{n}
\fi

\newcommand{\be}{\begin{equation}}
\newcommand{\ee}{\end{equation}}
\newcommand{\ba}{\begin{eqnarray}}
\newcommand{\ea}{\end{eqnarray}}
\newcommand{\baa}{\begin{eqnarray*}}
\newcommand{\eaa}{\end{eqnarray*}}
\begin{document}
  
\title{Weak Ferromagnetism and Excitonic Condensates}
\author{M.Y.~Veillette and L.~Balents}
\address{Physics Department, University of California, Santa Barbara, CA 93106}
\date{Date:\today}

\ifrevtex
  \maketitle  
\fi


\begin{abstract}
We investigate a model of excitonic ordering (i.e electron-hole pair condensation) appropriate for the divalent hexaborides. We show that the inclusion of imperfectly nested electron hole Fermi surfaces can lead to the formation of an undoped excitonic {\em metal} phase. In addition, we find that weak ferromagnetism with compensated moments arises as a result of gapless excitations.  We study the effect of the low lying excitations on the density of states, Fermi surface topology and optical conductivity and compare to available experimental data.\\
\\
PACS: 71.35.Cc, 71.35.-y, 71.20-b
\end{abstract}

\pacs{PACS: 71.35.Cc, 71.35.-y, 71.20-b}
\ifrevtex
\else
  \maketitle  
\fi
\begin{multicols}{2}
\section{Introduction}
The remarkable  discovery of weak ferromagnetism in lightly doped divalent hexaborides has been hailed as a great surprise in the physics of magnetism \cite{Young1,Rice1,Jarlborg,Varma1,Rodriguez,Gorkov1,Ichinomiya}. In spite of the absence of partially filled d- or f- orbitals, which are usually required for magnetism, these materials undergo a ferromagnetic transition at a relatively high Curie temperature (600K) and with a small magnetic moment in a narrow range of doping. Zhitomirsky {\em et al.} \cite{Rice1} have put forward the proposal of a doped excitonic insulator to explain the unexpected ferromagnetism in these materials, a theory that has a long and rich history \cite{Rice2}, but for which hexaborides seem to be the first experimental realization. Band structure calculations \cite{Massida} of $DB_6$ where $D= La, Ba, Sr$ indicate a small band overlap or small band gap at the three X points in the Brillouin zone, a condition favorable to the formation of an exciton condensate. The Coulomb attraction between particles and holes can overcome the small band gap and lead to the spontaneous condensation of electron-hole pairs. As a result of Hunds rule, a triplet exciton condensate is energetically favored over a singlet, giving rise to a local spin polarization within the unit cell but no net magnetization. 
The connection of exciton condensation to ferromagnetism was first pointed out by Volkov {\em et al.} \cite{Volkov1}.
Upon doping, a spin singlet exciton is formed in presence of the triplet order, breaking translational as well as time-reversal symmetry, and hence resulting in ferromagnetism. In more physical terms, the extra carriers added into the system are distributed asymmetrically between the two spin projections to preserve the most favorable pairing condition for one spin orientation.

Although appealing, the excitonic model has been largely studied only for highly idealized cases.  In this paper, we relax some of these approximations to extract generic predictions applicable to excitonic condensates. 
In particular, we are interested in explaining recent angle resolved photo emission spectroscopy (ARPES) \cite{Denlinger} measurements and de Haas-van Alphen (dHvA)  \cite{Hall} measurements on the Fermi surface in $Ca_{1-x}La_xB_6$ and $SrB_6$.  We study in some detail the zero temperature phase diagram as a function of the doping level in the presence of imperfectly nested electron and hole Fermi surfaces. 
We show a possible new phase, the excitonic metal (EM), can reconcile the existence of a Fermi surface and an excitonic condensate. Furthermore, as an upshot of the EM, the ungapped Fermi surface leads to a compensated magnetic moment,  a mechanism for partial polarization markedly different from the one proposed by Zhitomirsky {\em et al}. 

The paper is structured as follows. First, in the next Section we introduce a two-band model Hamiltonian and discuss its physical relevance to the excitonic state.  In Section III, we analyze the model for the case of perfectly nested Fermi surfaces. Using the techniques of mean-field theory, we recast the problem into a form closely related to BCS theory. We show that excitonic states with collinear spin polarization are favored in the ferromagnetic phase. In Section IV, the effect of incommensurability of the Fermi surface on the emergence of ferromagnetism is studied. We report numerical results on the magnetic moment and density of states in the excitonic phases. The effect of the excitonic order  on the optical conductivity is considered in Section IV. 

\section{Model}

As argued by Balents {\em et al} \cite{Varma1}, the binding of electron and hole is most effective for states with similar group velocity at equal momenta. Hence, in the hexaborides, the electron-hole pair will be more strongly bound if both components are drawn from the same X point of the Brillouin zone.
 Thus, we neglect the three-pocket band structure appropriate to the hexaborides to study a two band model of conduction and valence electrons.
To analyze the problem of excitonic ferromagnetism, we introduce the Green's function defined by the equality
\begin{equation}
{\cal G}^{\sigma \sigma^\prime}( {\bf r}, {\bf r^\prime}, \tau)= - \langle T_{\tau} \psi_\sigma({\bf r},\tau) \psi_{\sigma^\prime}^\dagger ( {\bf r^\prime},0) \rangle.
\end{equation}
Here $\psi_{\sigma}({\bf r},\tau)$ and $\psi^\dagger_{\sigma}({\bf r},\tau)$ are operators annihilating and creating an electron with spin $\sigma$ ($=\pm$) at point ${\bf r}$ at imaginary time $\tau$.  In the two band model, we can express the field $\psi^\dagger_\sigma$ in terms of $a^\dagger_{k,\sigma}$ and $b^\dagger_{k,\sigma}$, the creation operator in the conduction and valence band, respectively:
\begin{eqnarray}
\psi_{\sigma} ( {\bf r}) &=& \sum_k  \varphi_{a {\bf k}} ({\bf r}) a_{k,\sigma} + \varphi_{b {\bf k}} ({\bf r}) b_{k,\sigma}, \nonumber \\
\psi^\dagger_{{\sigma}} ( {\bf r}) &=& \sum_k  \varphi^\ast_{a {\bf k}} ({\bf r}) a^\dagger_{k,{\sigma}} + \varphi^*_{b {\bf k}} ({\bf r}) b^\dagger_{k,{\sigma}},
\end{eqnarray}
where $\varphi_{a {\bf k}}({\bf r})$ and $\varphi_{b {\bf k}}({\bf r})$ are the Bloch functions with quasimomentum $\hbar k$ of electrons in the conduction and valence bands, respectively. The Hamiltonian of the system can be written as $H=H_0 + H_I$, where
\begin{equation}
H_0= \sum_{k, \sigma} (\epsilon^a_k - \mu ) a^\dagger_{k,\sigma} a^{\phantom{\dagger}}_{k,\sigma}+(\epsilon^b_k-\mu)  b^\dagger_{k, \sigma} b^{\phantom{\dagger}}_{k,\sigma},
\end{equation}
is the non-interacting part of the Hamiltonian. $\epsilon^a_k$ and $\epsilon^b_k$ are the kinetic energy of conduction and valence electrons, respectively. The interacting part of the Hamiltonian is given by the Coulomb interaction
\begin{equation}
H_I= \frac{1}{2} \sum_{\sigma, \sigma^\prime} \int d {\bf r} d {\bf r^\prime} \psi^\dagger_{\sigma}({\bf r}) \psi^{\phantom{\dagger}}_{\sigma}({\bf r}) V({\bf r-r^\prime}) \psi^\dagger_{\sigma^\prime}({\bf r^\prime}) \psi^{\phantom{\dagger}}_{\sigma^\prime}({\bf r^\prime}),
\end{equation} 
where $V({\bf r})= \frac{e^2}{\epsilon} \frac{1}{|{\bf r}|}$.
The exciton Bohr radius $a_B$ sets the length scale of the problem, the physics being controlled by momentum exchange processes of the order of $1/a_B$. In the weak binding limit, i.e when the energy of an exciton $E_B$ is much smaller than the bandwidth $E_W$, $a_B$ is much larger than the interlattice spacing $a$ and the interaction is dominated by the long range part of the Coulomb interaction. In this case, we can treat the interaction in the so-called dominant term approximation \cite{Rice1}, $H_{DT}=H_0 + \overline{H_I}$, where
\begin{eqnarray}
\overline{H_I}=\frac{1}{2} \sum_{k,k^\prime,q,\sigma,\sigma^\prime} V({\bf q})
( a^\dagger_{k^\prime, \sigma^\prime} a^{\phantom{\dagger}}_{k^\prime -q, \sigma^\prime}+  b^\dagger_{k^\prime,\sigma^\prime}b^{\phantom{\dagger}}_{k^\prime-q, \sigma^\prime}) \times  \nonumber \\
(a^\dagger_{k, \sigma} a^{\phantom{\dagger}}_{k +q, \sigma}+  b^\dagger_{k,\sigma}b^{\phantom{\dagger}}_{k+q, \sigma}).
\end{eqnarray}
Here  $V({\bf q})= \frac{4 \pi e^2}{\epsilon  q^2}$ is the long range part of the Coulomb interaction. 
Notice that unlike the full crystal Hamiltonian $H$, $H_{DT}$
 conserves {\em separately} the charge and the spin of the conduction and valence electrons.  The order parameter $\langle a^\dagger_\sigma  b^{\phantom{\dagger}}_{\sigma^\prime}  \rangle$ that corresponds to the pairing of a valence and conduction electrons is a matrix in spin space.
Because the approximate Hamiltonian conserves separately the particle number of each species, the energy of the system is invariant under a constant shift of  the overall phase of the order parameter. Similarly, the $SU(2) \times SU(2)$ symmetry leads to the degeneracy of the excitonic triplet and singlet states.  The full crystal Hamiltonian will in general break these symmetries. 
However, we can justify the use of the approximate Hamiltonian by arguing the neglected terms will affect the formation of the excitonic condensate only at a very low energy scale.  Indeed,  these terms involve matrix elements of the Coulomb potential at wavevector $\sim 1/a$ and hence can be treated as small perturbations compared to the long range Coulomb interaction part of the interaction at wavevector $1/a_B$.  It is interesting nonetheless to study how they lift the degeneracy. There are three classes of terms that we have ignored. The first one contains  electron-hole exchange terms such as $a^\dagger_{k+q, \sigma} a^{\phantom{\dagger}}_{k^\prime, \sigma^\prime} b^\dagger_{k^\prime-q, \sigma^\prime} b^{\phantom{\dagger}}_{k ,\sigma}$.  These terms break the $SU(2) \times SU(2)$ symmetry explicitly. They are ultimately responsible for the splitting of the ground state of the Hamiltonian in favor of the triplet state \cite{Rice2}. Another class of terms left aside are matrix elements of the kind $a^\dagger_{k+q, \sigma} a^{\phantom{\dagger}}_{k, \sigma} b^\dagger_{k^\prime-q, \sigma^\prime} a^{\phantom{\dagger}}_{k^\prime, \sigma^\prime}$. In case of the hexaborides, these terms are prohibited at small $k, k^\prime, q$ due to lattice symmetry. The third class contains interband pair scattering terms that create and destroy simultaneously two valence and conduction electrons in the Brillouin zone. This term will select the {\em phase} of the order parameter. However, the effect of these virtual processes are limited by the low density of excited pairs. 

\section{Mean-Field Approach}

\begin{figure}[ht]
\begin{center}
 \epsfxsize=10.0cm
 \epsfysize=8.0cm
\rotatebox{0}{ \epsffile{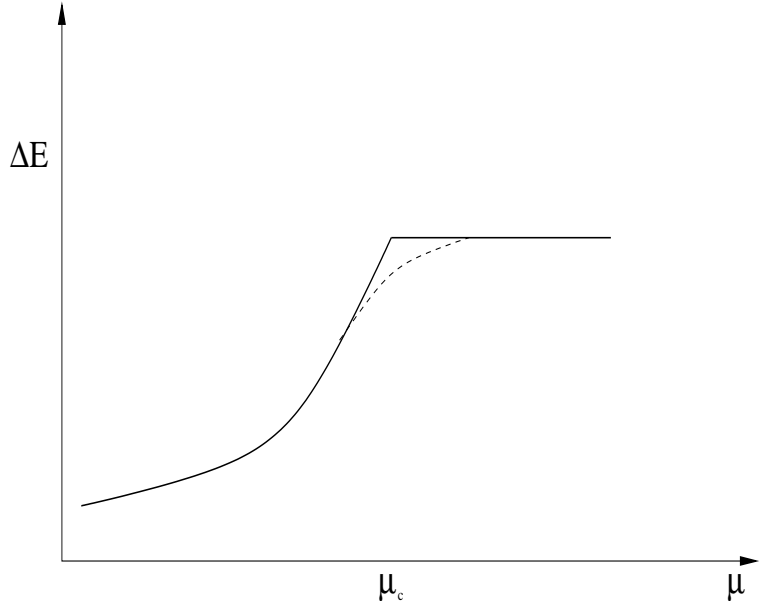}}
\end{center}
\vspace{0.2in}
\caption{Maxwell construction for the condensation energy:  In bold line, we plot equation \ref{condensation} for the condensation energy per spin species.  For collinear order parameter, the energy gain $\Delta E(\mu+h)+\Delta E(\mu-h)$ near the first order transition is {\em linear} in $h$ rather than quadratic. Shown in dash line is the gain in condensation energy resulting from the exchange term. 
}
\label{maxwell}
\end{figure}

First, we study the Hamiltonian $H_{DT}$ in a mean-field (MF) treatment.  For large exciton radius $a_B$, the excitons overlap strongly and we expect MF theory to give correct results. The excitonic phase is characterized  by the pairing of valence and conduction electrons. To emphasize the similarity of our treatment to the Nambu-Gorkov formalism for superconductors, we introduce the Green's function in the Bloch basis
\begin{displaymath}
{\cal G}^{{\sigma} {\sigma^\prime}}({\bf k}, \tau ) = - 
\left( 
\matrix{
\langle T_{\tau} a_{k,{\sigma}} (\tau) a_{k,{\sigma^\prime}}^\dagger (0) \rangle &
\langle T_{\tau} a_{k,{\sigma}} (\tau) b_{k,{\sigma^\prime}}^\dagger (0) \rangle \cr
\langle T_{\tau} b_{k,{\sigma}} (\tau) a_{k,{\sigma^\prime}}^\dagger (0) \rangle &
\langle T_{\tau} b_{k,{\sigma}} (\tau) b_{k,{\sigma^\prime}}^\dagger (0) \rangle \cr
}
\right).
\end{displaymath}
The excitonic gap $\Delta_{{\sigma} {\sigma^\prime}}$ is related to the off-diagonal Green's function 
\begin{equation}
\Delta_{{\sigma} {\sigma^\prime}}({\bf k}) =  \sum_q V({\bf k-q}) {\cal G}_{a b}^{{\sigma} {\sigma^\prime}}({\bf q},\tau=0^{-}).
\label{gap}
\end{equation}
Ignoring for the moment the possibility of ferromagnetic ordering, the MF Green's function is given by 
\begin{eqnarray}
{\cal G}^{{\sigma} {\sigma^\prime}}({\bf k}, i \omega )
&=& - \Big( -i \omega+g_k -\mu +\epsilon_k \tau^z \nonumber \\
&+&  \frac{ \Delta_{\sigma \sigma^\prime}}{2} (\tau^x -i\tau^y)+ \frac{\Delta^*_{\sigma^\prime \sigma}}{2} (\tau^x +i \tau^y) \Big)^{-1},
\label{MF1}
\end{eqnarray}
where $\epsilon_k= (\epsilon^a_k- \epsilon^b_k)/2$ and $g_k= (\epsilon^a_k+ \epsilon^b_k)/2$ and $\vec{\tau}$ are Pauli matrices in the band space.
 Eqn. \ref{MF1} must be solved in conjunction with the gap equation (Eqn. \ref{gap}) to obtain a self-consistent solution. At this point, it is convenient to introduce the decomposition of $\Delta_{{\sigma} {\sigma^\prime}}$ in terms of the singlet $\Delta_s$ and triplet $\vec{\Delta}_t$ order parameter.
\begin{equation}
\Delta_{\sigma \sigma^\prime}= \left( \Delta_s {\cal I} + \vec{\Delta}_t \cdot \vec{\sigma} \right)_{\sigma \sigma^\prime},
\end{equation}
where ${\cal I}$ and $\vec{\sigma}$ are respectively the $2 \times 2$ unit and Pauli matrices in the spin space. Because the Hamiltonian $H_{DT}$ is invariant under separate rotations of the valence and conduction electrons, the triplet states are degenerate with the singlet states. Consider the case where the order parameter is collinear with the spin, i.e $\vec{\Delta}_t = \Delta_t \hat{z}$. With this replacement, the MF Hamiltonian decouples the spin up and spin down electrons. It is straightforward to obtain the one-particle excitation spectrum of spin ${\sigma} (= \pm)$ electrons
\begin{equation}
\xi_{k {\sigma}}^{\pm}= -\mu +g_k \pm \sqrt{\epsilon_k^2+ \Delta^*_{\sigma} \Delta_{\sigma}},
\label{eigen}
\end{equation}
where $\Delta_{\sigma}= \Delta_s +{\sigma} \Delta_t $. In the theory of excitonic insulator, it is often assumed that the electronic dispersion relation are nested and isotropic, i.e. $\epsilon^a_k= - \epsilon^b_k$. As made apparent in equation \ref{MF1}, the particle-hole transformation $b_k \rightarrow b_{-k}^\dagger$ maps the problem onto two copies of the BCS model in a Zeeman field (a copy for each spin polarization). The Zeeman field takes the role of the chemical potential. This problem has been worked out, within a MF treatment, by Larkin {\em et al.} \cite{Larkin} and Fulde {\em et al.} \cite{Fulde}. Assuming strong screening of the Coulomb interaction, i.e. $V({\bf k}) \approx V_0$, we see from equation $\ref{gap}$, that the gap equation becomes momentum independent. At stoichiometry, we obtain a BCS-like state with a gap $|\Delta_{\sigma} |= \Delta_0 \equiv 2t \exp (-1/ N_0 V_0)$  where $t$ is a cutoff energy around the Fermi surface and $N_0$ is the density of states per spin at the Fermi level. The lower band $\xi^{-}$ is fully occupied whereas the upper band $\xi^{+}$ is empty: the system is an excitonic insulator with a one-particle excitation gap of $\Delta_{\sigma}$.  Turning on the chemical potential, the hole and electron Fermi surfaces shift with respect to one another. This loss of nesting lowers the condensation energy and ultimately the excitonic state disappears. For a uniform order parameter, the transition occurs at  $\mu=\mu_c \equiv \Delta_0 / \sqrt{2}$ via a first order transition. The condensation energy per spin direction relative to the normal state for $\mu<\mu_c$, is 
\begin{equation}
\Delta E_{\sigma}(\mu)= N_0 \left( \mu^2 - \frac{\Delta_0^2}{2} \right).
\label{condensation}
\end{equation}
At $\mu=\mu_c$, the ground state is degenerate and  {\em each} spin polarization can be in a state where $\Delta= 0$ or  $\Delta=\Delta_0$. Although the ground state has generally a 2-fold degeneracy at a first order transition point, the degeneracy for {\em each} spin species is non-generic. Up to this point, the MF treatment has neglected the effect of the intraband Coulomb interaction. While it is appropriate to ignore the effect of the direct part of the Coulomb interaction for uniform states, its exchange part yields a ferromagnetic interaction. The spin of the electrons in the conduction and valence band, respectively $\vec{s}_a=a^\dagger \vec{\frac{\sigma}{2}} a$  and $\vec{s}_b=b^\dagger \vec{\frac{\sigma}{2}} b$, interact via the Heisenberg Hamiltonian $H_S= -J \left(\vec{s}^2_a + \vec{s}^2_b \right) $ where $J>0$. In our mean-field theory, $J= 2V_0$. However, the exchange constant tends to be poorly approximated by the direct coupling constant because the true exchange is screened by the dielectric function of the crystal. Hence we take $J$ to be an adjustable parameter interaction (although still of order $V_0$). Phenomenologically, $J$ is related to the antisymmetric Landau-Fermi liquid parameter $F_0^A$ via the relation $J= -\frac{F_0^A}{N_0}$ \cite{Abrikosov}. Notice that there are additional terms in the full crystal Hamiltonian that couple to the excitonic order parameter but are negligibly small compare to $V_0$ and hence can be safely ignored in first approximation. 
To take into account the exchange energy, we introduce a generalized mean-field theory, replacing $-J \vec{s}_a^2  \rightarrow  -2 \vec{h}_a \cdot \vec{s}_a + \frac{\vec{h}_a^2}{J}$ and similarly for the valence electrons. We find 
\begin{eqnarray}
{\cal G}({\bf k}, i \omega )
&=& - \Big( -i \omega+g_k -\mu +\epsilon_k \tau^z \nonumber \\
&+&  \frac{ \Delta_s +\vec{\Delta} \cdot \vec{\sigma}}{2} (\tau^x -i\tau^y)+ \frac{\Delta_s^* + \vec{\Delta}^* \cdot \vec{\sigma}}{2} (\tau^x +i \tau^y) \nonumber \\
&-& (\vec{h}_a+\vec{h}_b)  \cdot \vec{\sigma}   
-(\vec{h}_a-\vec{h}_b) \cdot \vec{\sigma} \tau^z  \Big)^{-1},
\label{MF2}
\end{eqnarray}
where the self-consistent equations are
\begin{eqnarray}
\vec{h}_a=  \frac{J}{2} \sum_{k} Tr\left[ \vec{\sigma} {\cal G}_{aa}({\bf k},\tau=0^-)  \right], \nonumber \\
\vec{h}_b=  \frac{J}{2} \sum_{k} Tr\left[ \vec{\sigma} {\cal G}_{bb}({\bf k},\tau=0^-)  \right].
\end{eqnarray}
We have studied the energetics for collinear spin polarization and perfectly nested Fermi surfaces. While the spin branches are no longer decoupled, we find the condensation energy of the system $\Delta E_T$ is given by a simple Maxwell construction
\begin{equation}
\Delta E_{T}(\mu)= \Delta E_{+}(\mu+h) + \Delta E_{-}(\mu- h) +\frac{2 h^2}{J},
\label{Maxwell}
\end{equation}
where $h=| \vec{h}_a|= | \vec{h}_b|$. Minimizing this functional, we find that an excitonic ferromagnet ground state is favored in a range of chemical potential $\mu_c \sqrt{1-2 N_0 J} < \mu < \mu_c \sqrt{\frac{1-2N_0 J}{1-4N_0 J}}$ (see Fig. \ref{maxwell}). This excitonic polarized state is characterized by having one spin species, say spin up, in the normal state whereas the spin down electrons are paired  ($| \Delta_{-}|=\Delta_0$). One can understand the origin of the magnetism by studying its effect on the Fermi surfaces. The magnetic field favors the pairing of the spin down electrons by increasing the  coincidence of the hole and electron Fermi surfaces for the spin down. This gain in condensate energy is {\em linear} in $h$ near the first order transition.  While this redistribution of the charge carriers comes at the expense of a loss in kinetic energy for the spin up electrons, this increase in energy is quadratic in the field therefore favoring ferromagnetic ordering. 

We have investigated the energetics of order parameters non-collinear with the spin. An instance of such of state is $\Delta_{\sigma \sigma^\prime}=\frac{1}{2} (\sigma_{x}+i\sigma_{y}) \Delta_0$. This particular state has been proposed by Zhitomirsky {\em et al.} \cite{Rice1} to account for ferromagnetism in the hexaborides. Unlike in the collinear states, the ferromagnetic moment,
\begin{equation}
\vec{m}=  \frac{1}{2} \sum_k Tr[ \vec{\sigma} {\cal G}({\bf k}, \tau=0^- )],
\end{equation}
is perpendicular to the exciton polarization $\vec{m} \perp \vec{\Delta_t}$. For non-collinear spin polarization, the simple Eqn.\ref{Maxwell} is no longer valid. Instead, we find that the gain in condensate energy in the ferromagnetic non-collinear state is  {\em quadratic} in $h$, so it is never the lowest energy state. An inquiring reader might object that such a state could lower its energy by taking advantage of terms that we ignore in the dominant-term approximation. While the energy difference of the non-collinear and collinear state is of the order of  $( N_0 \Delta_0^2) N_0 J$, the correction brought by the neglected terms are of the order $N_0 \Delta^2 (N_0 \delta V)$ where $\delta V \sim V_0 (a/a_B)^2$. In strong coupling where $a_B \sim a$, one needs to reconsider the problem and it was found that the excitonic order in the ferromagnetic regime may be of non-collinear type  \cite{Leon1}.
Physically, the key signature of the non-collinear state $\frac{1}{2} (\sigma_{x}+i\sigma_{y}) \Delta_0$  is the spontaneous generation of spin currents, markedly absent in collinear spin polarized states. However, it is known to be notoriously  difficult to observe spin currents experimentally. A possible way to distinguish the two phases is by studying their collectives modes although this warrant further investigations. 

\section{Incommensurability}
\begin{figure}[ht]
\centerline{\epsfxsize=4.0cm
\epsfysize=4.0cm
\rotatebox{0}{\epsffile{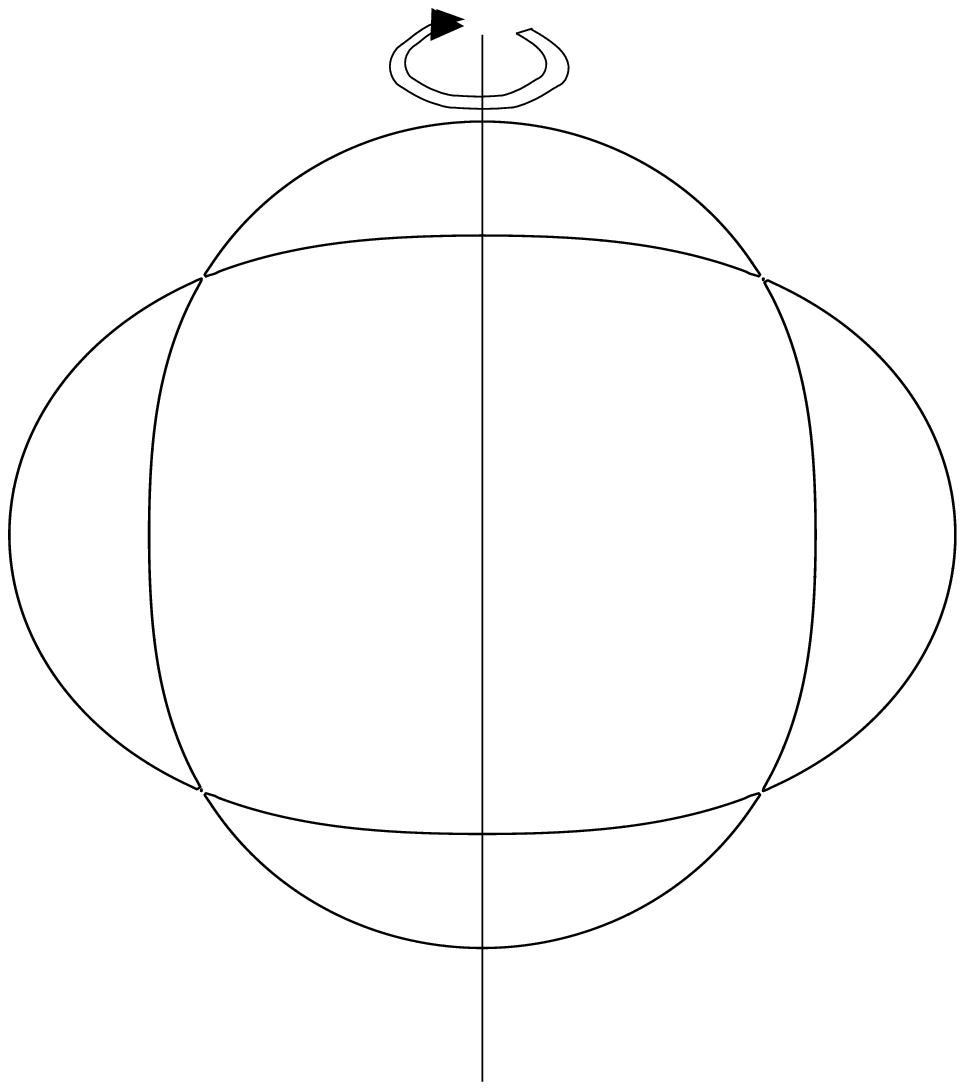}} 
(A)
\epsfxsize=4.0cm
\epsfysize=4.0cm
\rotatebox{90}{\epsffile{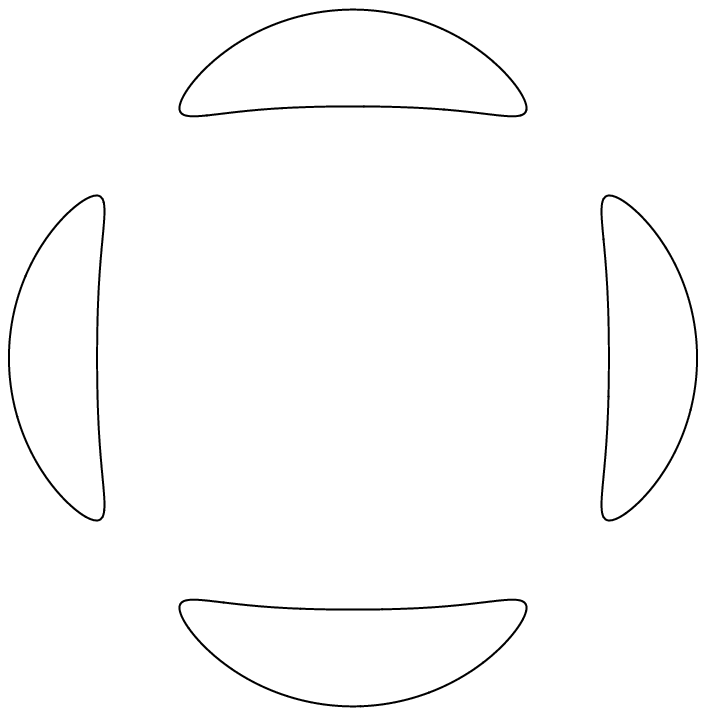}}
(B)
\epsfxsize=4.0cm
\epsfysize=4.0cm
\rotatebox{90}{\epsffile{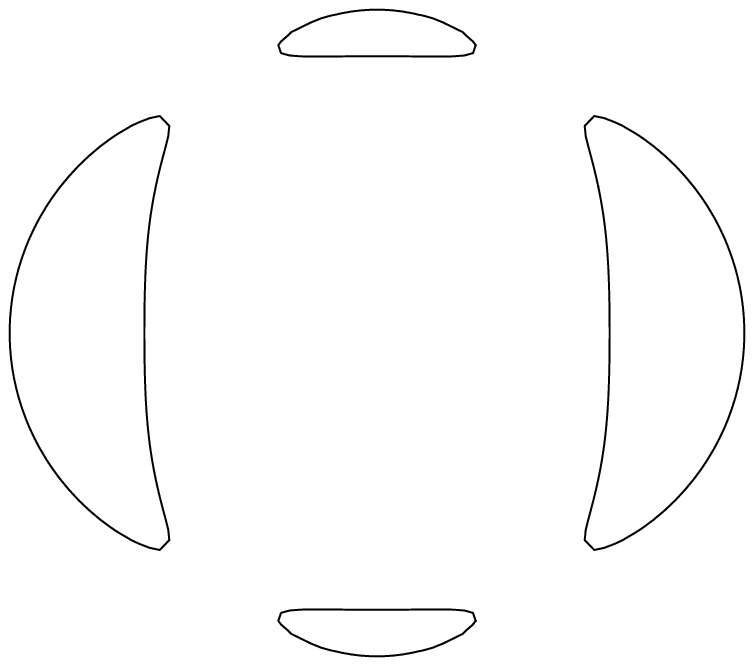}} 
(C)
}
\vspace{0.2in}
\caption{Fermi surface cut through the z-x plane where the  z-axis is an axis of revolution. For non-interacting electrons, the imperfectly nested Fermi surface consists in two ellipsoids (A). In the undoped EM phase (B), the Fermi surface is partially gapped, forming a hole lens and an electron ring. Upon doping (C), the hole lens drops out leaving an enlarged electron ring.}
\label{FS}
\end{figure}

Although the previous results are suggestive, a complete understanding of the ferromagnetism and its applicability to the hexaborides is still lacking. 
The large number of compounds exhibiting weak ferromagnetism discovered to this day argues for the robustness of the phenomenon. However, most of theoretical  results seemingly hinge on the perfect nesting of the hole-electron Fermi surface to lead to an excitonic insulator. Moreover, the existence of a Fermi surface as seen in ARPES and dHvA data rules out this simple scenario.  A crucial step to circumvent these difficulties is to realize the hole and electron bands are not related by any symmetry principle and in real materials, electron and hole Fermi surfaces are not generally perfectly nested. To study the effect of incommensurability on the excitonic states,  we introduce the anisotropic dispersion relations for the valence and conduction bands 
\begin{eqnarray}
\epsilon_k^{a}&=& v(|k|-k_F)  + \delta  v k_F \cos (2\theta) - \mu^\star,   \nonumber \\
\epsilon_k^{b}&=&-v(|k|-k_F) + \delta  v k_F \cos (2\theta) - \mu^\star. 
\end{eqnarray}
Here $v$ is the Fermi velocity, $k_F$ is the Fermi wavevector, $\delta$ is a dimensionless parameter that controls the degree of antinesting and $\theta$ is the azimuthal angle. For $\delta=0$, we recover the isotropic and nested dispersion relations  whereas for finite $\delta$ the Fermi surface is described by two non-coinciding ellipsoids (See Fig. \ref{FS}).  The parameter $\mu^\star$ is determined by imposing charge neutrality at stoichiometry at $T=0$. i.e.
\begin{equation}
\sum_k \left[ \Theta(-\epsilon^a_k)+\Theta(-\epsilon^b_k) \right]= \sum_k 1,
\end{equation}
where $\Theta(x)$ is the Heaviside function. This toy model characterization of the energy dispersion describes simply the effect of anisotropy as well as allows us to make analytic progress. Furthermore, the large uncertainties in the available data for the effective masses and gap values did not allow for an accurate theoretical description although LDA calculations \cite{Rice1} on $CaB_6$ would suggest $\delta$ to be positive.  

\begin{figure}[ht]
\begin{center}
 \epsffile{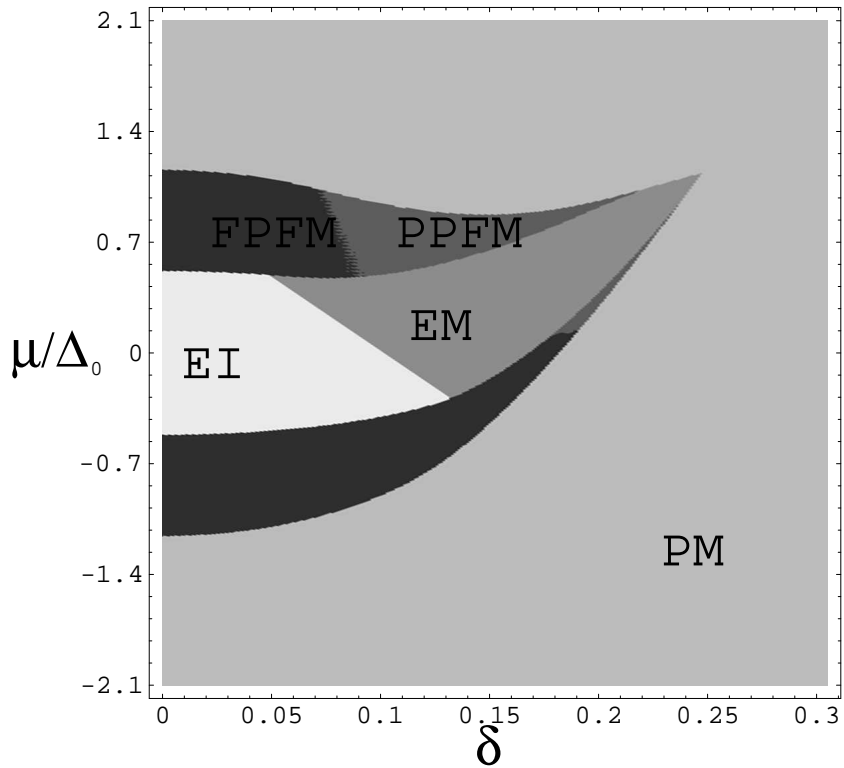}
 \epsfxsize=10.0cm
 \epsfysize=8.0cm
\end{center}
\vspace{0.2in}
\caption{Phase Diagram as a function of the chemical potential $\mu$ and the antinesting parameter $\delta$. PM indicates a paramagnetic metallic phase, EI and EM are the excitonic insulating and metallic phase, respectively. FPFM and PPFM denote the fully polarized excitonic ferromagnet and the partially polarized excitonic ferromagnet, respectively. Most phase transitions are continuous. The exceptions are the PPFM/FPFM and the EI/EM boundaries.} 
\label{pd}
\end{figure}

\begin{figure}[ht]
\begin{center}
 \epsfxsize=10.0cm
 \epsfysize=8.0cm
 \epsffile{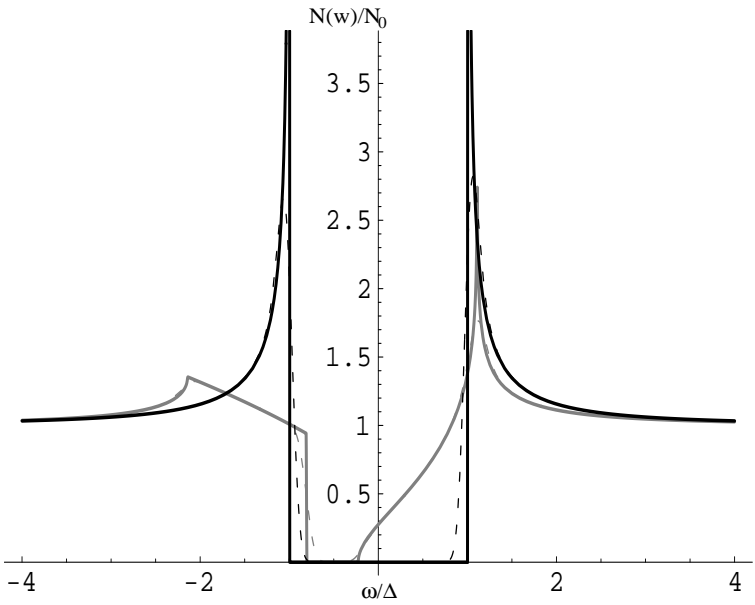}
\end{center}
\vspace{0.2in}
\caption{Electronic Density of States (DOS): The DOS in the excitonic insulator and metal phase indicated in black and gray line, respectively. The dash lines represent the DOS in presence of scattering for values of  $\gamma/ \Delta=0.1$.(See text) }
\label{dos}
\end{figure}

Based upon our previous analysis, we consider only collinear states and minimize the energy functional
\begin{eqnarray}
E_T[\mu]=  \sum_{\alpha,\sigma=\pm} \left[ \int \frac{d^3 k}{(2 \pi)^3} (\xi^\alpha_{k,\sigma}- \sigma h) \Theta(-\xi^\alpha_{k,\sigma}+\sigma h) \right] \nonumber \\
+ \frac{2 h^2}{J}+ \frac{\Delta_+^2+\Delta_-^2 }{V_0}, 
\end{eqnarray}
where $\xi^\alpha_{k,\sigma}$ is the energy eigenvalue determined in Eqn.  \ref{eigen}. The integral over wavevector can be expressed in terms of elliptical functions. From the three parameters $(h, \Delta_{+}, \Delta_{-})$ minimization search, we have determined the phase diagram at zero temperature as a function of $\mu$, the chemical potential and $\delta$, the antinesting parameter (See Fig.\ref{pd}). Not surprisingly, for small values of $\delta$, we find that the system behaves analogously to the nested case. At stoichiometry, the system is in the excitonic insulating (EI) phase. Increasing the chemical potential, the EI state becomes unstable to a ferromagnetic phase before disappearing into the normal state via a first order transition. However, larger values of $\delta$ allows for the possibility of gapless excitations in the one particle excitation spectrum.  Hence, while the excitonic metal (EM) exhibits a non-zero order parameter i.e. ($\Delta_s=0, \Delta_t \neq 0$ or $\Delta_t=0, \Delta_s \neq 0$), the presence of an incompletely gapped Fermi surface modifies radically its electronic properties.  To illustrate this point, we have calculated the electronic density of states (DOS), 
\begin{equation}
N(\omega)=-\frac{1}{2 \pi} Im \sum_k Tr [{\cal G}({\bf k}, \omega)],
\end{equation}
at stoichiometry in the insulating and metallic phases.  As shown in Fig. \ref{dos}, in spite of the excitonic ordering, the DOS in the metallic phase does not show the typical shoulder edge present in the insulating excitonic phase. The EM/EI transition is continuous and the excitonic gap varies smoothly across the transition line defined by $\mu=\mu_t$. However, the metallic instability is signaled by the density of states at the Fermi level which show a singular behavior across the transition line, increasing as $\sqrt{\mu-\mu_t}$.  The introduction of charge carriers in the EM phase raises the Fermi level, leading to changes in the topology of the Fermi surface. Recent results on dHvA experiments on divalent hexaborides have shown the presence of Fermi sheets in the undoped compounds \cite{Hall}. The Fermi surface is thought to consist of two pieces, an electron ring and a hole lens centered around the X-point in the Brillouin Zone. In the La-doped $CaB_6$, dHvA data show the hole pocket dropping out, yielding a single electron sheet. These results are consistent with our model. 
As seen in Fig. \ref{FS}, in the undoped case, the Fermi surface is gapped at the intersection of the hole and electron Fermi surface, leaving two Fermi sheets, a ring and a lens. For positive values of $\delta$, the lens and ring are  hole-like and electron-like, respectively. As electrons are added to the system, the hole pocket shrinks and eventually becomes gapped out due to the excitonic gap, leaving a single electron ring, consistent with experimental data. However, further experiments on cleaner samples are needed to clearly determine the topology of the Fermi surface.

An upshot of the finite density of states at the Fermi level is the possibility of distributing the extra carriers asymmetrically between spin species. As shown in Fig.\ref{gap}, the EM phase gives way to a {\em partially} polarized ferromagnet where the extra carriers carry only a {\em fraction} of a Bohr magneton. When $\delta$ becomes of the order of $\Delta/ \epsilon_F$, the incommensurability of the hole and electron Fermi surfaces becomes too large and hinders the gap formation. However, this state may be itself unstable to the formation of non-homogeneous states with spin and/or charge density waves\cite{Gorkov1}. 

\begin{figure}[ht]
\begin{center}
 \epsfxsize=10.0cm
 \epsfysize=8.0cm
\rotatebox{0}{\epsffile{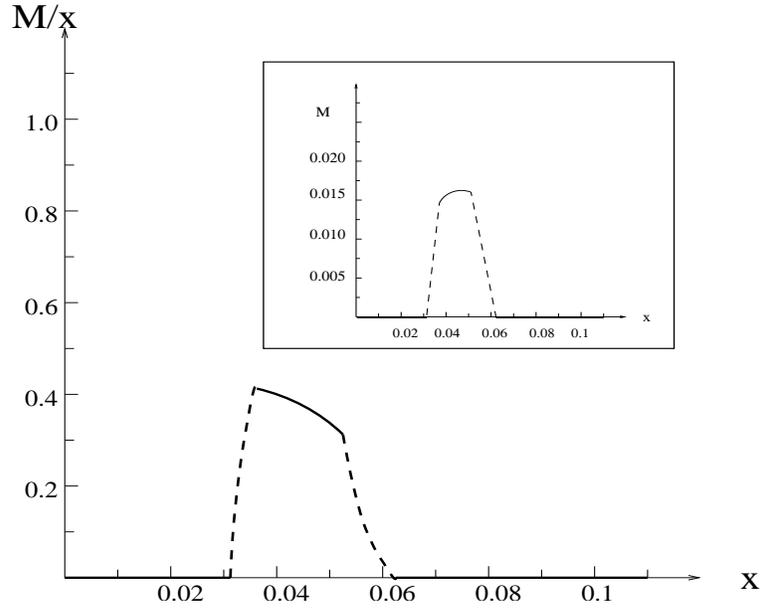}}
\end{center}
\vspace{0.2in}
\caption{Magnetic moment in Bohr magneton per unit carrier as a function of doping at T=0 for $\delta=0.15$. In the phase separation regime, represented by the dashed line, we expect the magnetization density to vary linearly between the two competing phases as shown in the inset.}
\label{mm}
\end{figure}

The result of our calculation of the magnetic moment per dopant for the doped excitonic metal is shown in Fig.\ref{mm}. In contrast to the case of a doped excitonic insulator, the ferromagnetic moment per carrier is smaller than a Bohr magneton due to the electrons spilling over both branches of the spin species.
Unlike in the EI/FPFM transition, the electronic charge density can in principle evolve smoothly across the EM/PPFM boundary, thus one might expect a second order (continuous) transition to arise. However, we find in MFT a first order transition. Moreover, we suspect this may be a general result due to singular quasiparticle-mediated interactions. An argument in favor of a first order transition is found by considering the system at low but non-zero temperature. In the neighborhood of the transition, were it second order, it should be possible to develop a Landau expansion of the free energy in terms of the magnetic order parameter $h$ with $\Delta$ already non-zero. This expansion must be analytic at $T>0$
\begin{equation}
F(\mu,h)= F(\mu,0) + A h^2 + B h^4 + ....
\end{equation}
The parameters A and B are related to two- and four-point spin correlation functions respectively. As $T \rightarrow 0$, we find $A= \frac{2}{J}-4 N_0(\mu)$ and $B \rightarrow -\infty$ hence precluding the validity of such an expansion and therefore of a continuous transition. It would be desirable to develop a proper field theoretic argument posed directly at $T=0$. A consequence of the direct first-order transition is the discontinuous jump of the electronic density across the boundary line. Since experiments are performed at fixed charge density, phase separation occurs for a range of doping.  However, the long range Coulomb interaction will likely frustrate macroscopic phase separation and an inhomogeneous state is expected.

\section{Electromagnetic Absorption}

\begin{figure}[ht]
\begin{center}
 \epsfxsize=13.0cm
 \epsfysize=8.0cm
 \epsffile{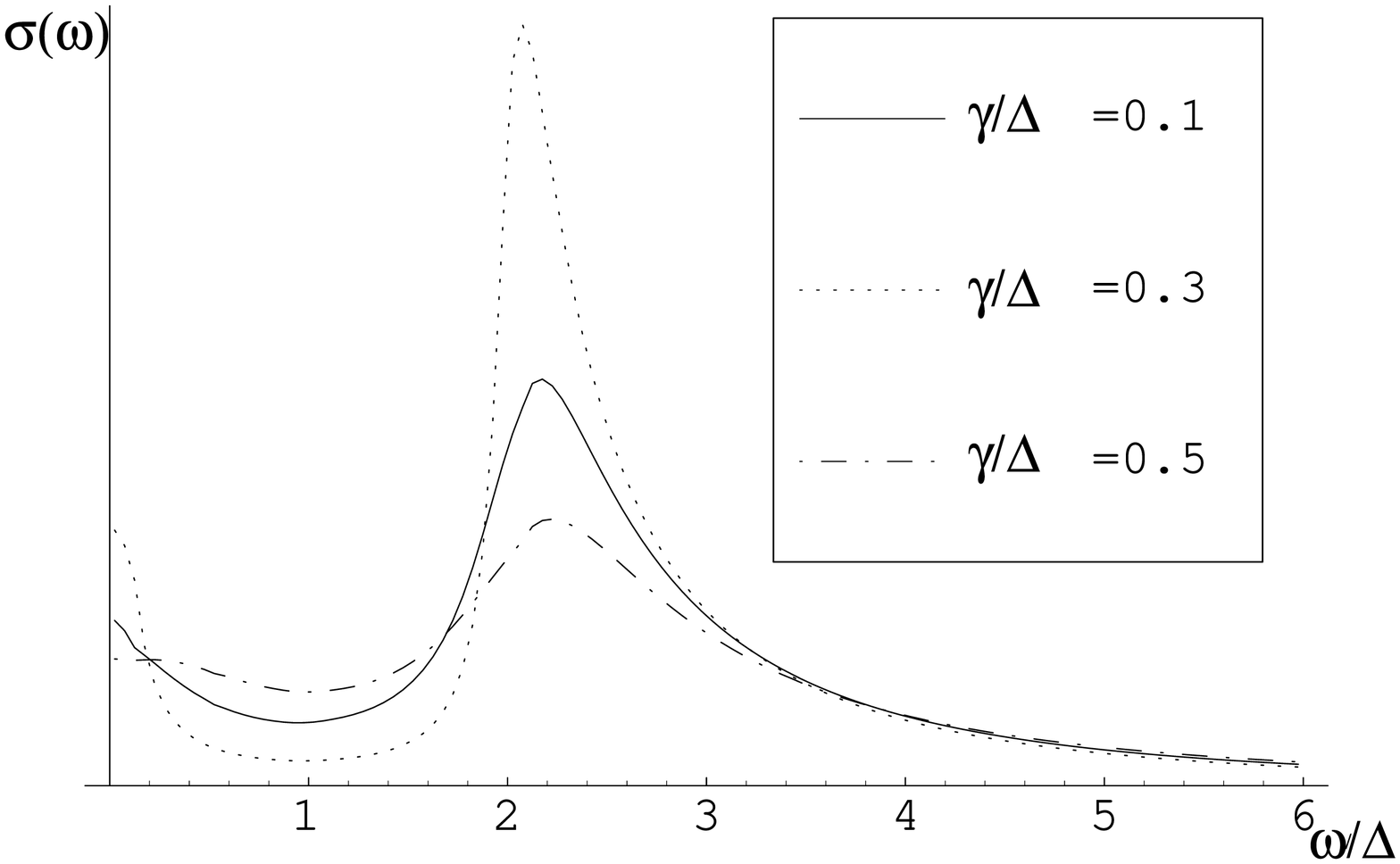}
\end{center}
\vspace{0.2in}
\caption{ Electromagnetic absorption as a function of frequency $\omega$ in the EM phase. The optical edge near $2 \Delta$ is smeared by scattering processes. }
\label{abs}
\end{figure}

A powerful technique that can be used to probe the excitonic state is the infrared optical conductivity. Just as for superconductors, the excitonic insulator signature on the optical conductivity is the suppression of absorption for frequencies twice below the excitonic gap $\Delta_0$. However, one of the puzzling aspects of optical conductivity measurements on the undoped hexaborides is the absence of hard optical gap and strong doping dependence \cite{Vonlanthen}. One expects the presence of gapless quasiparticles in the excitonic metal state to be responsible for this result. To investigate this issue quantitatively  we calculate the optical conductivity tensor $\sigma_{\mu \nu}(\omega)$ which is related to the current-current correlation function $\Pi_{\mu \nu}$
\begin{equation}
\sigma_{\mu \nu}(\omega>0)= \frac{-1}{\omega} Im  \int_{-\infty}^{\infty} dt  e^{i \omega t} \Pi_{\mu \nu}(t),
\end{equation}
where  $\Pi_{\mu \nu}(t)= -i \Theta(t) \langle [J_\mu (t), J_\nu (0)] \rangle $ and $\vec{J}$ is the current operator. The Greek indices refer to the spatial components. We calculate the correlation function in the Matsubara formalism and obtain the desired retarded function by analytic continuation.  To allow for electronic scattering due to impurities and/or lattice imperfections, we introduce a finite lifetime to the quasiparticle by performing the substitution  $\omega \rightarrow \omega +\gamma sign(\omega)$ in 
 the Green's function
\begin{equation}
{\cal G}^{\sigma \sigma^\prime} ({\bf k}, i \omega)=  \frac{\delta_{\sigma \sigma^\prime}}{i \omega +i\gamma sign(\omega) -\mu +g_k+ \epsilon_k \tau^z +\Delta_\sigma \tau^x},
\end{equation}
where for concreteness, we consider a pure triplet order parameter. 
Ignoring vertex corrections to the conductivity, the correlation function is given by the bubble diagram, 
\begin{eqnarray}
\Pi_{\mu \nu}(i \omega)= e^2 \int  \frac{d\Omega}{2\pi} \frac{d^3k}{(2\pi)^3}  Tr \left[ {\cal G}({\bf k},i\Omega-i\omega/2) \right. \nonumber \\
\left. v_\mu {\cal G} ({\bf k},i\Omega+i\omega/2)  v_\nu \right] ,
\end{eqnarray}
where $v_\mu = \partial_\mu(\epsilon_k \tau^z - g_k)$ is the velocity of the quasiparticles. In order to evaluate the correlation function, we use the standard spectral representation for the Green's function 
\end{multicols}
\begin{equation}
{\cal G}^{\sigma \sigma^\prime}({\bf k}, i \omega)= \int \frac{d\rho}{2\pi}  \frac{1}{\rho-i \omega} 
2 Im \left[\frac{\delta_{\sigma \sigma^\prime}}{\rho+i\gamma -\mu +g_k+ \epsilon_k \tau^z +\Delta_\sigma \tau^x} \right]. 
\end{equation}
\begin{multicols}{2}
In the physical regime where the scattering rate $\gamma$ is much smaller than $\Delta$, the imaginary part of the correlation function in real frequencies is given by 
\end{multicols}
\begin{eqnarray}
Im[\Pi_{\mu \nu}(\omega)]=\sum_{\sigma} \delta_{\mu \nu} \frac{(e v_F)^2}{3} \int  \frac{d^3k}{(2\pi)^3}  \frac{2 \gamma}{\omega^2+ \gamma^2} \frac{1}{4 \chi^2 +\gamma^2} \Big[ \left( \Theta(\xi^{+}_{\sigma})- \Theta(\xi^{+}_{\sigma}+\omega )-\Theta(\xi^{-}_{\sigma})+ \Theta(\xi^{-}_\sigma -\omega ) \right)  \nonumber \\
\frac{(2 \chi^2)(\chi^2+ (\chi+\omega)^2) +4 \chi (\chi+\omega)(\chi^2-2\Delta^2)}{(\omega+2\chi)^2+\gamma^2} - (\omega \rightarrow -\omega) \Big],
\end{eqnarray} 
\begin{multicols}{2}
\noindent where $\chi=\sqrt{\epsilon_k^2+\Delta^2}$. We evaluate the correlation in different limits. For low frequencies, i.e. $\omega \ll \Delta$, we find a Drude-like conductivity,
\begin{equation}
\sigma_{\mu \nu}(\omega)= \delta_{\mu \nu} \frac{(e v_F)^2 \gamma}{\omega^2 +\gamma^2} n,
\end{equation}
where
\begin{equation}
n=\frac{1}{3} \sum_{\sigma=\pm} \sum_{\alpha=\pm} \int \frac{d^3k}{(2\pi)^3}  \frac{\epsilon_k^2}{\epsilon_k^2+\Delta_\sigma^2}  \delta(\xi^\alpha_{k,\sigma}).
\end{equation}
Hence, the scattering of gapless excitations generates absorption in the low energy sector of the optical conductivity. For high frequencies, the dominant contribution to the integral comes from exciton pair breaking, i.e. when $\omega \simeq 2 \chi$. Thus for $\omega \ll \gamma$
\end{multicols}
\begin{equation}
\sigma_{\mu \nu}(\omega)= 4 \delta_{\mu \nu}  \sum_{\sigma} \frac{(e v_F \Delta_\sigma)^2}{\omega^3} \int \frac{d^3k}{(2\pi)^3} \frac{2 \gamma}{(\omega-2\chi)^2+\gamma^2}  \left[\Theta(\xi^{+}_{k,\sigma})-\Theta(\xi^{+}_{k,\sigma})\right].
\end{equation}
\begin{multicols}{2}
\noindent In the clean limit, the threshold for exciton pair breaking is given by $ 2\Delta$. In this case the absorption peak shows a square root singularity $ \sim \sqrt{\frac{1}{\omega^2-4 \Delta^2}}$ near the threshold due to the large amount of phase space available in $k$ space. However, in  presence of scattering, the wavevector $k$ is no longer a good quantum number and the peak is smeared out on a width of order $\gamma$ (see Fig \ref{abs}). We conjecture that the absence of hard optical gaps in experimental data is due to metallicity  and scattering that shifts the absorption to lower frequencies. Notice that because the order parameter for the spin up and spin down is generally different in the ferromagnetic state, in clean materials, the absorption peak would be split into two peaks upon entering the ferromagnetic phase. Its experimental observation in the hexaboride materials would provide considerable support for the theory of excitonic ferromagnet.  

\section{conclusion}

Excitonic ordering near a semiconductor-metal transition offers a natural explanation of weak ferromagnetism in the doped hexaborides.  We have generalized previous studies on the excitonic ferromagnet by considering the effect of imperfect nesting on the excitonic state. Adding this significant ingredient, we showed that a novel phase, the excitonic metal, is stable and can account for the Fermi surface seen in the hexaborides. A ferromagnetic instability still occurs in the excitonic metal near the first order transition. Furthermore, the small ferromagnetic moment observed in lightly doped hexaborides is naturally explained as a result of compensated moments on the imperfectly gapped Fermi surface.  Although this model qualitatively explains experiments, further studies will be needed to determine the effect of the gapless excitations on the electronic transport and thermodynamic properties of the system. 

\section{acknowledgements}

L.B. and M.Y.V. were supported by the NSF--DMR--9985255 and the Sloan
and Packard foundations.


\end{multicols}

\end{document}